# TAXON version 1.1: A simple way to generate uniform and fractionally weighted three-item matrices from various kinds of biological data

Evgeny V. Mavrodiev, Florida Museum of Natural History, University of Florida, Gainesville, FL, 32611, USA, e-mail: evgeny@ufl.edu

Alexander Madorsky, Department of Physics, University of Florida, Gainesville, FL, 32611, USA, e-mail: madorsky@phys.ufl.edu

An open-access program allowing three-item statement matrices to be generated from data such as molecular sequences does not exist so far. The recently developed LisBeth package (ver. 1.0) allows representing hypotheses of homology among taxa or areas directly as rooted trees or as hierarchies; however, this is not a standard matrix-based platform. Here we present "TAXON version 1.1" (TAXON), a program designed for building three-item statement-matrices from binary, additive (ordered) and non-additive (unordered) multistate characters, with both fractional and uniform weighting of the resulted statements.

## 1. Introduction

Three-item analysis reduces information on taxon relationships to a series of three-item statements (3TSs) of the form A(BC): taxa B and C are related to each other more closely than to taxon A (Nelson and Platnick 1991). The 3TSs, even as implemented in conventional data matrices, are the high level hypothesis concerning the relationships of two taxa relative to a third, not low-level hypothesis about character state distribution within the standard matrix (Platnick et al. 1996). 3TS data therefore is an entirely different way of viewing information – representing relationships directly (Williams and Ebach 2008). However an open-access program allowing three-item statement matrices to be generated from data such as molecular sequences does not exist so far. The recently developed LisBeth package (ver. 1.0) allows representing hypotheses of homology among taxa or areas directly as rooted trees or as hierarchies (Ducasse et al. 2007), but this is not a standard matrix-based platform.

We present "TAXON version 1.1" (TAXON), a program designed to build 3TS-matrices from binary (**b**), additive (ordered) (**om**) and non-additive (unordered) multistate characters (**umc**) (up to 25 symbols), with the addition of the IUB/IUPAC codes (DNA/RNA/AAs) and with both fractional/uniform weighting of the resulting statements.

## 2. Implementation

TAXON has a simple command line interface described in Supplement 1. Utility is written in portable C++. The source code compiles equally well with Microsoft Visual

C++ 2010 Express and GNU C++ compiler version 3.4.4 under Cygwin. Porting this code to any other platform with a standard C++ compiler should be possible.

TAXON accepts files in Comma Separated Value (CSV) format as input. These files can be generated from programs such as Excel, OpenOffice Calc, etc. as well as Mesquite package (Maddison and Maddison 2011). Output data can be written in simplified NEXUS and PHYLIP formats, as well as CSV (Supplement 1, Supplementary examples 1-3).

Currently, the maximum number of taxa/characters in the input matrix must not exceed 5000/100000 respectively. These values can be modified within the source code if necessary (See Supplement 1 for details).

TAXON will be freely available upon request.

### 3. Discussion

The MS-DOS program TAX (incl. MATRIX and MOMATRIX) (Nelson and Ladiges 1994, reviewed in Williams and Siebert 2000) renders a matrix of standard characters into a matrix of three-item statements, creating appropriate output for analysis with parsimony programs such as Henning86 and NONA (reviewed in Williams and Siebert 2000). Nelson and Ladiges (1992), however, did not address the issue of **umc** (review in Williams and Siebert 2000) and the conversion of **umc**-matrices to series of 3TSs is the subject of discussion (e. g., Platnick 2009).

Nelson and Platnick (1991) mentioned that to extract all possible information from **umc** distributions it may be necessary to examine separately all possible orderings, the 3TSs that each implies. Later, Nelson and Ladiges (1992) and Williams and Ebach (2008) suggested that from the perspective of 3TS-analysis, a multi-state character is equivalent to a suite of 3TSs with no statements appearing more than once.

In TAXON we implemented what we believe to be the simplest way of converting a **umc** matrix to a 3TS matrix as the explication of all possible triplets of relationships **by exhaustion of the outgroup value**. Within the resulting 3TS matrix, each taxon therefore may be represented with all-possible "minimal" relationships with all other taxa of the same matrix. We call this method of umc data transformation to 3TSs a "general way" (G).

Williams and Siebert (2000, see also Nelson and Platnick 1991) pointed out that 3TS approach *a priori* estimates a putative synapomorphy or code the data relative to an *a priori* defined out-groups. Therefore another way to build a 3TS-matrix is the coding of standard data relative to the **fixed value of the outgroup**. This method of transforming may be called "Williams's" (**W**) (Williams and Siebert 2000: 194, Table 9.5). Both **G** and **W** methods do not require the initial non-additive re-coding of the standard **umc**-matrix (Carine and Scotland 1999).

If a matrix contains states 0/1 and state 0 *a priori* is assumed to be plesiomorphic, we call this matrix "binary" (Nelson and Platnick 1991). Without this latter assumption, 0/1 matrix is a particular case of **umc**-matrix.

Therefore, using both **G**/**W** methods we can re-build **umc**-matrix to 3TS-matrix in two ways: 1. we may still keep multistate notation with the resulted 3TS matrix, or 2. we may present 3TS matrix as binary matrix (**WS** conversion, see Williams and Siebert 2000). Both of these ways are implemented in TAXON (Supplementary example 2).

When analyzed, the binary/multistate-notated 3TS may provide the same or similar results using standard parsimony analysis, but not necessarily using other methods of phylogeny reconstructions.

TAXON converts both **b** and **om** matrices to 3TS-matrices as described in Nelson and Platnick (1991) (Supplementary example 1).

Fractional weighting (FW) (Nelson and Ladiges 1992, 1994), to compensate for the influence of putative redundant statements, is also implemented in TAXON, at this time withoutthe elimination statements from 3TS matrix (Supplementary example 3).

In the case of **G**-conversion of **umc** matrices formulas from Nelson and Ladiges (1992) have been modified accordingly:

$$N_{3TS} = \sum_{k=1}^{k}[(t - n_k)(n_k - 1)n_k]/2;$$

$$N_{i3TS} = \sum_{k=1}^{k}(t - n_k)(n_k - 1);$$

$$FW = N_{i3TS}/N_{3TS},$$

where $N_{3TS}$ is the total number of 3TSs, $N_{i3TS}$ is the number of independent 3TSs, *t* is a number of taxa, *n* is the number of taxa with the informative states, and *k* is a number of informative state of **umc**.

In case of **W**-conversion of **umc** matrices the weighting of the resulted 3TSs is the subject of future consideration. In recent version of TAXON we implemented the possibility to weigh (W) the final 3TSs by fraction:

$$W = N_{3TSW}/N_{3TS},$$

where $N_{3TSW}$ is the total number of 3TSs if the value of the Outgroup fixed.

In case of ambiguities, the average weight of 3TSs is assigned to the statement. In cases when all characters of a standard matrix contain ambiguities weighting is disabled.

As a starting point, we recommend to use G-converted/uniformly weighted/multistate notated 3TS-matrixes for Maximum Likelihood approach as well as for phenetic algorithms of clustering like neighbor joining, and W-converted/uniformly weighted 3TS-matrixes for standard parsimony. In cases of **umc**-matrices that help to prevent the effect of grouping simply by all similarities described by Kluge and Farris (1999) for some cases of 3TS-matrices generated from non-additive binary data.

## 4. Conclusions

As a proof of the potential of TAXON to build the 3TS matrices we presented the ML tree (Figure 1) built using the IUB-notated 3TS matrix generated from the standard matrix of phytochrome locus C (PHYC) obtained from Saarella et al. (2007, Supplemental Figure 2) following **G-**conversion.

## Acknowledgments

We thank Dr. D. Williams (London Museum of Natural History) for very helpful discussion and comments. No agreement implied.

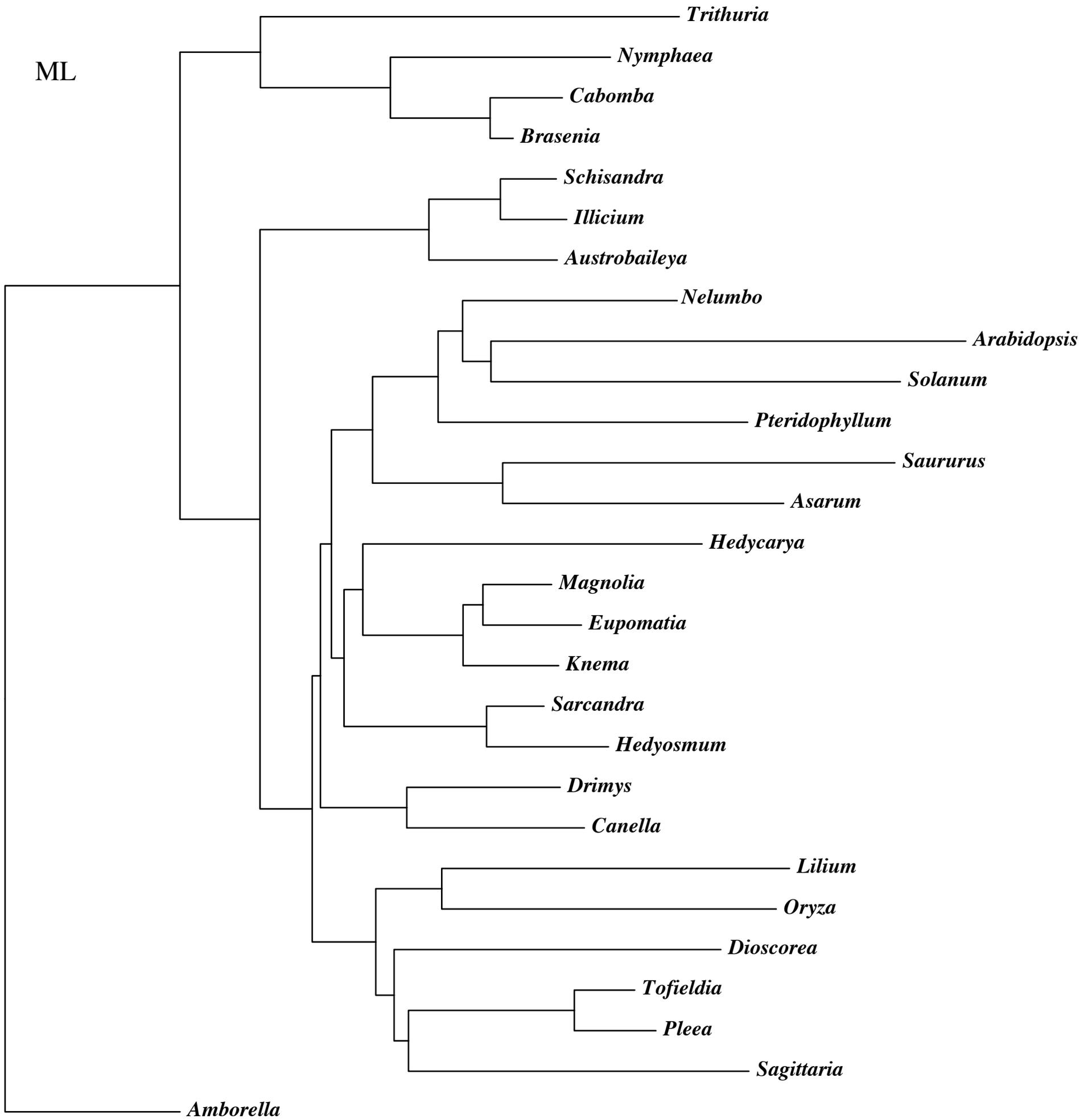

Figure Legend

Figure 1. Maximum Likelihood tree for the analysis of the IUB-notated 3TS matrix of phytochrome locus C (PHYC) (Log likelihood = – 4127677.799014 ). The standard PHYC matrix of 28 taxa and 1 228 characters obtained from Saarella et al. (2007, Supplementary Figure 2). All gaps/ambiguities treated as a missing data. G-conversion performed. The total number of characters in 3TS matrix is **1 078 731**, all variable. Analysis compiled using PhyML (Guindon, Gascuel 2003) as implemented in a SeaView (ver. 4.0)(Gouy et al. 2010). The GTR model assumed as the best choice and *Amborella trichopoda* selected as an Outgroup.

# Supplement 1

**TAXON ver. 1.1** does not require any installation process. Run the program without arguments via command-line reference. Utility is designed for building three-item statement (3TS)-matrices from binary, ordered and unordered multistate characters, with fractional and uniform weighting of the resulting statements.

**Command line interface**

First argument must always be the name of the CSV file with input matrix. One or several options in any order may follow the input file name. Table 1 shows the list of available options.

Table 1. TAXON v1.1: Options

| Option | Description |
|---|---|
| **Input symbols** | |
| -ib | input: binary (default) |
| -iom | input: ordered multistate |
| -ium | input: unordered multistate |
| -idna | input: DNA/RNA |
| -ip | input: protein |
| **Conversion method** | |
| -m3 | method: 3TS (default, G-conversion = the value of the outgroup exhaustive) |
| **Output symbols** | |
| -ob | output: binary (default) |
| -om | output: multistate |
| -odna | output: DNA/RNA |
| -op | output: protein |
| **Fractional weights** (**NEXUS files only**) | |
| -fw | print fractional weights and save all 3TSs in matrix (default: off) |
| **Outgroup** | |
| -og | print outgroup (default: off) |
| **Output formats** | |
| -phy | enable PHYLIP output (default: on if no other output selected) |
| -nex | enable NEXUS output (default: off) |
| -csv | enable CSV output (default: off) |
| | |

Table 2 TAXON v1.1:  Input file-symbols

| Input option | Symbols |
|---|---|
| Binary | 0 1 |
| Ordered multistate | 0 1 2 3 4 5 6 7 8 9 : < = > @ A B C D E F G H I J K |
| Unordered multistate | 0 1 2 3 4 5 6 7 8 9 : < = > @ A B C D E F G H I J K |
| DNA/RNA (IUB-codes) | A C G T U R Y S W K M B D H V |
| Protein (IUPA-codes) | A C D E F G H I K L M N P Q R S T V W Y |

**Example 1**

Input matrix format example is shown below:

```
taxonA,0,0

taxonB,=,0

taxonC,>,3

taxonD,@,4

taxonE,@,6
```

First (leftmost) column contains names of taxa, all following columns contain characters. Symbols allowed for each input option are shown in

**Table 2**.

**Additionally, input file can contain a pre-defined outgroup taxon name (W-conversion).** It must always be **last line** in the input file, in the following format:

```
Out,taxonB
```

In the example above, "Out" is a reserved keyword. No real taxa must be named with that name in user's input files. "taxonB" is the name of the outgroup taxon.

**Example 2**

G-conversion with the binary 3TS matrix output from standard DNA matrix in simplified NEXUS format with outgroup added, all 3TS fractionally weighted:

**`taxon.exe input.csv -idna -ob -og -fw -nex`**

**Please note that the command line interface may change in future versions.** Please see the documentation provided with each version of the utility for complete details.

**Limitations and performance**

Currently, the maximum count of taxa in the input matrix must not exceed 5000, and the maximum count of characters is 100000. These values can be modified in the source code if necessary. The output matrix is constructed entirely in computer's RAM before being written on disk. If a computer has enough RAM to accommodate the entire output matrix then the processing will occur with maximum possible performance. If the amount of RAM is not sufficient, a typical operating system (such as Windows or Linux) will attempt to use disk swapping. This will affect the performance severely, but the program will still finish processing. Finally, if the size of the disk swap file is not sufficient, TAXON will report memory allocation error and show the amount of memory required to accommodate the output matrix. In such case, the user should increase the size of the disk swap file and rerun the utility.

SUPPLEMENTARY EXAMPLE 1

I. Matrix 1 (after Nelson and Platnick, 1991: 354, Matrix 5).

```
A   00010
B   00101
C   01001
D   11111
E   10000
```

A. Output 1. File in CSV format:

```
    1,1,1,2,2,2,3,3,3,4,4,4,5,5,5,5,5
A   0,7,7,0,7,7,0,7,7,1,1,1,0,7,0,7,0,7
B   7,0,7,7,0,7,1,1,1,0,7,7,1,1,1,1,7,7
C   7,7,0,1,1,1,7,0,7,7,0,7,1,1,7,7,1,1
D   1,1,1,1,1,1,1,1,1,1,1,1,7,7,1,1,1,1
E   1,1,1,7,7,0,7,7,0,7,7,0,7,7,0,7,7,0
```

B. Output 2. File in simplified NEXUS format/Uniform Weighting:

```
#NEXUS
Begin DATA;
Dimensions ntax=6 nchar=18;
FORMAT SYMBOLS="0 1" MISSING=? GAP=-;
Matrix
A   077077077111070707
B   707707111077711177
C   770111707707117711
D   111111111111771111
E   111770770770707070
Out 000000000000000000
;
End;
```

C. Output 3. File in PHYLIP format:

```
6 18
A   077077077111070707
B   707707111077711177
C   770111707707117711
D   111111111111771111
E   111770770770707070
Out 000000000000000000
```

II. Character 1 (after Nelson and Platnick, 1991: 358).

```
A   0
B   1
C   1
D   2
E   3
F   3
```

Step 1: additive recoding of Character 1 (II):

```
A   0,0,0
B   1,0,0
C   1,0,0
D   1,1,0
E   1,1,1
F   1,1,1
```

Step 2: calculation of 3TSs based on Matrix, step 1:

1. Output 1. File in simplified NEXUS format/Uniform Weighting:

```
#NEXUS
Begin DATA;
Dimensions ntax=7 nchar=23;
FORMAT SYMBOLS="0 1" MISSING=? GAP=-;
Matrix
A   00000000007707707077
B   11117777770707077077
C   17771177777077077077
D   71777711711111777777
E   77277117111117711111
F   77727717177711111111
Out 0000000000000000000000
;
End;
```

2. Output 2. File in PHYLIP format:

```
7 23
A   00000000007707707077
B   11117777770707077077
C   17771177777077077077
D   71777711711111777777
E   77277117111117711111
F   77727717177711111111
Out 0000000000000000000000
```

3. Output 3. File in CSV format:

```
    1,1,1,1,1,1,1,1,1,2,2,2,2,2,2,2,3,3,3
A   0,0,0,0,0,0,0,0,0,7,7,0,7,7,0,7,7,0,7,7
B   1,1,1,1,2,7,7,7,7,7,0,7,7,0,7,7,0,7,0,7
C   1,7,7,1,1,7,7,7,7,7,0,7,7,0,7,7,0,7,0,7
D   7,1,7,7,7,1,1,1,1,1,1,1,1,7,7,7,7,7,7,0
E   7,7,1,7,7,1,7,7,1,1,1,7,1,1,1,1,1,1,1,1
F   7,7,7,1,7,7,1,1,7,7,1,1,1,1,1,1,1,1,1,1
Out 0,0,0,0,0,0,0,0,0,0,0,0,0,0,0,0,0,0,0,0
```

SUPPLEMENTARY EXAMPLE 2

Examples of conversion of the unordered (non-additive) multisite character to 3TS-equivalents

Character 1 (after Williams and Siebert, 2000: 194, Table 9.5).

A  0
B  1
C  1
D  2
E  2

1. G-conversion/Multistate Notation/Uniform Weighting

Output file in simplified NEXUS format:

```
#NEXUS
Begin DATA;
Dimensions ntax=6 nchar=6;
FORMAT SYMBOLS=" 0 1 2 3 4 5 6 7 8 9 : ; < = > # A B C D E F G H I J K" MISSING=? GAP=-;
Matrix
A    077077
B    111717
C    111771
D    727222
E    772222
Out  022011
;
End;
```

2. G-conversion/Binary Notation/Uniform weighting

Output file in simplified NEXUS format:

```
#NEXUS
Begin DATA;
Dimensions ntax=6 nchar=6;
FORMAT SYMBOLS=" 0 1" MISSING=? GAP=-;
Matrix
A    077077
B    111707
C    111770
D    707111
E    770111
Out  000000
;
End;
```

3. R-conversion/Outgroup fixed as a value of taxon A/Multistate Notation/Uniform Weighting

Output file in simplified NEXUS format:

```
#NEXUS
Begin DATA;
Dimensions ntax=6 nchar=2;
FORMAT SYMBOLS=" 0 1 2 3 4 5 6 7 8 9 : ; < = > # A B C D E F G H I J K" MISSING=? GAP=-;
Matrix
A    00
B    17
C    17
D    72
E    72
Out  00
;
End;
```

4. R-conversion/Outgroup fixed as value of taxon A/Binary Notation/Uniform Weighting (NS-conversion, proposed by Williams and Siebert, 2000: 193-195, Table 9.5).

Output in simplified NEXUS format:

```
#NEXUS
Begin DATA;
Dimensions ntax=6 nchar=2;
FORMAT SYMBOLS=" 0 1" MISSING=? GAP=-;
Matrix
A    00
B    17
C    17
D    71
E    71
Out  00
;
End;
```

SUPPLEMENTARY EXAMPLE 3

Fractional weighting (FW) as implemented in TAXON ver. 1.1

I. Character 1 (after Nelson and Ladiges, 1992: 492-493, Tab. 3, Fig. 3).

A    0
B    1
C    2
D    2

Output in simplified NEXUS format:

```
#NEXUS
Begin DATA;
Dimensions ntax=5 nchar=5;
FORMAT SYMBOLS="0 1" MISSING=? GAP=-;
Matrix
A    00007
B    11770
C    17111
D    71111
Out 00000
;
End;
Begin paup;
weights 0.667: 1, 0.667: 2, 0.667: 3, 1: 4, 1: 5;
End;
```

Note, that all 3TSs saved in 3TS-matrix.

If no 3TSs appearing more than once, fractional weights of the same 3TSs are added, and the "extra" 3TSs removed from the 3TS-matrix (Nelson and Ladiges, 1992: 492-493, Tab. 3, Fig. 3; see also Williams and Siebert, 2000: 190-193, Table 9.3 and Goloboff and Nixon, 1998 as reviewed in Nelson and Platnick, 1991: 355).

In last case, the NEXUS output file corresponding to Character 1 (I) should be:

```
#NEXUS
Begin DATA;
Dimensions ntax=5 nchar=4;
FORMAT SYMBOLS="0 1" MISSING=? GAP=-;
Matrix
A    0007
B    1170
C    1711
D    7111
Out 0000
;
End;
Begin paup;
weights 0.667: 1, 0.667: 2, 1.667: 3, 1: 4;
End;
```

This way of weighting of 3TSs will be implemented in a future versions of TAXON.

II. Character 2 (after Williams and Siebert, 2000: 194, Table 9.5).

A    0
B    1
C    1
D    2
E    2

1. G-conversion/Binary Notation/Fractional Weighting

Output file in simplified NEXUS format:

```
#NEXUS
Begin DATA;
Dimensions ntax=6 nchar=6;
FORMAT SYMBOLS="0 1" MISSING=? GAP=-;
Matrix
A    077077
B    111707
C    111770
D    707111
E    770111
Out 000000
;
End;
Begin paup;
weights 1: 1, 1: 2, 1: 3, 1: 4, 1: 5, 1: 6;
End;
```

2. H-conversion/Binary Notation

Output in simplified NEXUS format:

```
#NEXUS
Begin DATA;
Dimensions ntax=6 nchar=2;
FORMAT SYMBOLS="0 1" MISSING=? GAP=-;
Matrix
A    00
B    17
C    17
D    71
E    71
Out 00
;
End;
Begin paup;
weights 0.333: 1, 0.333: 2;
End;
```